\documentclass[aps,prl,reprint,superscriptaddress]{revtex4-2}

\usepackage{bm}
\usepackage{xcolor}
\usepackage{amsmath}
\usepackage{amsfonts}
\usepackage{amssymb}
\usepackage{natbib}
\usepackage{babel}
\usepackage{latexsym}
\usepackage{float}
\usepackage{verbatim}
\usepackage{bbold}
\usepackage{pbsi}
\usepackage[T1]{fontenc}
\usepackage{graphicx}
\usepackage{hyperref}
\hypersetup{colorlinks = true,
	linkcolor = blue,
	anchorcolor = blue,
	citecolor = blue,
	filecolor = blue,
	urlcolor = blue}

\begin{document}

\title{Frozen-In Gravitational Fields}

\author{Felipe A. Asenjo}
\email{felipe.asenjo@uai.cl}
\affiliation{Facultad de Ingenier\'ia y Ciencias,
Universidad Adolfo Ib\'a\~nez, Santiago 7491169, Chile.}
\author{Maricarmen A. Winkler}
\email{maricarmen.winkler@edu.uai.cl} 
\affiliation{Facultad de Ingenier\'ia y Ciencias,
Universidad Adolfo Ib\'a\~nez, Santiago 7491169, Chile.}
\author{Luca Comisso}
\email{luca.comisso@columbia.edu}
\affiliation{Department of Astronomy, Columbia University, New York, NY 10027, USA} 
\affiliation{Department of Physics, Columbia University, New York, NY 10027, USA}
\affiliation{Columbia Astrophysics Laboratory, Columbia University, New York, NY 10027, USA}

\begin{abstract}
Spacetime can undergo complex nonlinear evolution, as governed by the Einstein field equations, and a central challenge is to understand the geometric structures that arise, persist, and interact throughout its evolution. Using a formulation of Einstein equations that parallels nonlinear electrodynamics of continuous media, we show that general relativity admits gravitational field connections---two-surfaces and associated field lines whose connectivity is maintained by the spacetime dynamics. This gravitational frozen-in behavior is enabled by an ideal Ohm-type condition for the gravitational field. We further show that the same framework naturally leads to a conserved ``gravitational magnetic'' flux. A conserved gravitational helicity also emerges, with a clear topological interpretation in terms of gravitational field-line structures. These results identify well-defined topological constraints on admissible spacetime evolution and provide an organizing principle underlying the nonlinear dynamics of spacetime.
\end{abstract}

\maketitle

\textit{Introduction.}---
The evolution of spacetime curvature as governed by the Einstein field equations underlies some of the most energetic phenomena in the universe, including black-hole and neutron-star mergers, core-collapse supernovae, and the formation of large-scale cosmological structure. Understanding the spacetime evolution in a way that is both predictive and physically transparent remains a central challenge in general relativity. Progress in modeling such highly dynamical spacetimes has relied heavily on numerical relativity simulations \cite{Lehner01,Alcubierre08,BS10}. These simulations have played a crucial role in modeling compact-object coalescences \cite{Pretorius2005,Campanelli2006,Baker2006} and in interpreting gravitational-wave observations \cite{Abbott2016,Abbott2017}, and they remain an essential tool for exploring the nonlinear predictions of the Einstein field equations.

Alongside numerical simulations, a complementary and equally fundamental approach seeks to understand spacetime evolution in terms of its underlying geometric structure. This viewpoint, often referred to as geometrodynamics \cite{Wheeler63,Wheeler64}, has long been recognized as a powerful route to understanding spacetime dynamics \cite{MTW,Thorne12}. A key development has been the realization that the Einstein equations admit reformulations in terms of quantities that resemble electric and magnetic fields \citep[e.g.][]{MaarBass98}, through which spacetime curvature is organized into field-like geometric structures that provide an intuitive description of its dynamics. This approach has led to specific applications, including the use of tendex and vortex lines to visualize tidal gravitational accelerations and differential frame dragging in strong-gravity regimes \cite{Owen11,Nichols11,Nichols12,TB17}. More recently, a Maxwell-type interpretation of nonlinear gravity has provided new insight into the dynamical spacetime of binary-black-hole collisions and gravitational-wave turbulence \cite{boye,holly}. These studies showed that electrodynamical formulations of general relativity offer a powerful way to analyze the nonlinear dynamics of spacetime. 

In this Letter, we show that viewing gravity through an electrodynamical formulation reveals a precise connectivity structure of the gravitational field. Spacetime dynamics admits gravitational field connections—two-surfaces and associated field lines whose connectivity is preserved under evolution by the Einstein equations. This preservation arises because the gravitational field is transported by a velocity field satisfying an ideal Ohm-type condition, in analogy with ideal magnetohydrodynamics. Within the same framework, we demonstrate the conservation of a ``gravitational magnetic'' flux and identify a conserved gravitational helicity with a direct topological interpretation in terms of twist, writhe, and linkage of field lines. These results place topological constraints on admissible spacetime evolution and provide a geometric framework for understanding the nonlinear dynamics of spacetime. 
Throughout this Letter, we adopt units with $G=c=1$, and the metric signature $(-,+,+,+)$.

\textit{Theoretical Framework.}---
For the purposes of this Letter, it is enough to recall the dynamical equations of general relativity, which follow from the Einstein-Hilbert action 
\begin{equation}
S_{\rm EH} = \int d^{4}x \, \sqrt{-g} \, R \, ,
\end{equation}
supplemented by a matter action, whose variation with respect to the spacetime metric $g^{\mu\nu}$ yields 
\begin{equation}
G_{\mu\nu} = 8\pi T_{\mu\nu} \, ,
\label{eq:EFE}
\end{equation}
with $G_{\mu\nu}$ the Einstein tensor and $T_{\mu\nu}$ the matter stress-energy tensor.
To recast these equations into a form that shares formal structural features with a Yang-Mills-type formulation, thereby providing a gauge-theoretic framework suited to our analysis of the connectivity properties of the gravitational field, it is convenient to introduce a local tetrad $\mathcal{A}^{\hat{\alpha}}{}_\mu$ \cite{maluf23} related to the metric by 
\begin{equation}
\mathcal{A}^\mu{}_{\hat{\alpha}} \, \mathcal{A}^\nu{}_{\hat{\beta}} \, g_{\mu\nu} = \eta_{\hat{\alpha}\hat{\beta}}  \, ,
\end{equation}
with $\eta_{\hat{\alpha}\hat{\beta}}$ the Minkowski metric. Greek indices denote spacetime components, while hatted indices refer to the locally inertial tetrad. Covariance under local Lorentz transformations requires a spin connection $\omega^{\hat{\beta}}{}_{\hat{\alpha}\mu}$, that generalizes the covariant derivative for general tensors $V^\nu{}_{\hat{\alpha}}$ as $D_\mu V^\nu{}_{\hat{\alpha}} = \partial_\mu V^\nu{}_{\hat{\alpha}} 
+ \Gamma^\nu{}_{\mu\kappa} V^\kappa{}_{\hat{\alpha}}
+ \omega^{\hat{\beta}}{}_{\hat{\alpha}\mu} V^\nu{}_{\hat{\beta}}$, where $\Gamma^\nu{}_{\mu\kappa}$ is the Levi-Civita connection. 
A quantity formally analogous to an electromagnetic field strength can be constructed from the antisymmetric derivative of the tetrad, 
\begin{equation}
\mathcal{F}^{\hat{\alpha}}{}_{\mu\nu} = \partial_\mu \mathcal{A}^{\hat{\alpha}}{}_\nu - \partial_\nu \mathcal{A}^{\hat{\alpha}}{}_\mu \, ,
\label{eq:field_strength_tensor}
\end{equation} 
which defines a field strength in a tetrad-based representation of the gravitational field, with the tetrad viewed as a gauge potential. This gravitational field-strength tensor serves as the basic tensorial element entering the spin connection and the Maxwell-like rewriting of the Einstein equations \cite{olivares,Peshkov}. Projections along a timelike unit vector $n_\mu$ then define the associated ``gravitational electric'' and ``gravitational magnetic'' fields \cite{Lichnerowicz67,Anile05}
\begin{align}
\mathcal{E}^{\hat{\alpha}\mu} &= n_\lambda\,\mathcal{F}^{\hat{\alpha}\mu\lambda} , \label{eq:E_tensor}\\
\mathcal{B}^{\hat{\alpha}\mu} &= n_\lambda\,{^*\mathcal{F}^{\hat{\alpha}\mu\lambda}} , \label{eq:B_tensor}
\end{align} 
where ${^*\mathcal{F}}^{\hat{\alpha}}{}_{\mu\nu} = \frac{1}{2}\,\epsilon_{\mu\nu\rho\sigma}\, \mathcal{F}^{\hat{\alpha}\rho\sigma}$ is the dual field-strength tensor. Metric compatibility determines the spin connection in terms of the gravitational field strength via 
\begin{align}
\omega_{\hat{\alpha}\hat{\beta}\hat{\gamma}}
= \frac{1}{2} \left( \mathcal{F}_{\hat{\beta}\hat{\alpha}  \hat{\gamma}} + \mathcal{F}_{\hat{\gamma}\hat{\alpha}  \hat{\beta}} - \mathcal{F}_{\hat{\alpha}\hat{\beta}\hat{\gamma}}\right) \, , 
\end{align}
and from its antisymmetric and trace parts one can construct the dual Nester-Witten form \cite{Nester81,Witten81}, 
\begin{align}
^*{\mathcal{U}}^{\hat{\beta}\hat{\gamma}}{}_{\hat{\alpha}} = \omega^{[\hat{\beta}\hat{\gamma}]}{}_{\hat{\alpha}} + \delta^{\hat{\beta}}{}_{\hat{\alpha}} \omega^{[\hat{\gamma}\hat{\delta}]}{}_{\hat{\delta}} - \delta^{\hat{\gamma}}{}_{\hat{\alpha}} \omega^{[\hat{\beta}\hat{\delta}]}{}_{\hat{\delta}} \, .
\end{align}
This object allows the tetrad-projected Einstein tensor to be represented as a covariant derivative term plus a self-interaction term \cite{Frauendiener90,Szabados92}, 
\begin{equation}
G^{\mu}{}_{\hat{\alpha}}
= \nabla_{\nu} {^*{\mathcal{U}}_{\hat{\alpha}}}{}^{\nu\mu} - t^{\mu}{}_{\hat{\alpha}} \, ,
\label{eqn:G_NW}
\end{equation} 
where the Sparling self-current $t^{\mu}{}_{\hat{\alpha}}$ is given by 
\begin{equation}
{t^{\hat\mu}}_{\hat\nu}={\mathcal{F}^{\hat\alpha}}_{\hat\beta\hat\nu}\, {{^*{\cal U}}_{\hat\alpha}}^{\hat\beta\hat\mu}- \frac{1}{4} {\delta^{\hat\mu}}_{\hat\nu} {\mathcal{F}^{\hat\alpha}}_{\hat\beta\hat\lambda}\, {{^*{\cal U}}_{\hat\alpha}}^{\hat\beta\hat\lambda} \, .
\label{eqn:selfcurrent}
\end{equation}
The Sparling-Nester-Witten decomposition, \eqref{eqn:G_NW}-\eqref{eqn:selfcurrent}, then allows the Einstein field equations \eqref{eq:EFE} to be written in the form 
\begin{equation}
\nabla_\mu\,  {^*{\cal U}_{\hat\alpha}}^{\mu\nu} = {t_{\hat\alpha}}^\nu + 8\pi \, {T_{\hat\alpha}}^\nu \, ,
\label{einsteinEqs2}
\end{equation}
\begin{equation}
\nabla_\mu\,  {{^*\mathcal{F}}_{\hat\alpha}}^{\mu\nu}=0\, .
\label{homogeneous}
\end{equation}
Because ${^*{\cal U}_{\hat\alpha}}^{\mu\nu}$ is antisymmetric in its spacetime indices, these equations imply the conservation law $\nabla_\nu ({t_{\hat\alpha}}^\nu + 8\pi \, {T_{\hat\alpha}}^\nu)=0$. 
The Einstein equations thus take on a structure that resembles that of nonlinear electrodynamics of continuous media \cite{LL_V8}, which proves convenient for our analysis.

\textit{Gravitational Field Connections.}---
To examine the conditions under which the gravitational field preserves its connectivity, we consider the evolution of the gravitational field-strength tensor ${\mathcal{F}^{\hat\alpha}}_{\mu\nu}$. Writing Eq.~\eqref{homogeneous} in the form of a Bianchi-type identity, $\nabla_\lambda {\mathcal{F}^{\hat\alpha}}_{\mu\nu}+\nabla_\mu {\mathcal{F}^{\hat\alpha}}_{\nu\lambda}+\nabla_\nu {\mathcal{F}^{\hat\alpha}}_{\lambda\mu}=0$, its contraction with a general timelike vector field $u^\lambda$ gives 
\begin{eqnarray}
    u^\lambda \nabla_\lambda{\mathcal{F}^{\hat\alpha}}_{\mu\nu}&=&-u^\lambda  \nabla_\mu {\mathcal{F}^{\hat\alpha}}_{\nu\lambda}+u^\lambda \nabla_\nu {\mathcal{F}^{\hat\alpha}}_{\lambda\mu}\, .
    \label{connThe1}
\end{eqnarray} 
A key structural assumption, mirroring the ideal-conductivity condition in the electrodynamics of continuous media, is to invoke the condition 
\begin{equation}
    u^\lambda {\mathcal{F}^{\hat\alpha}}_{\lambda\nu}=0\, .
    \label{connThe0}
\end{equation}
For a tetrad adapted to $u^\lambda$ and Fermi-Walker transported along it, this condition can be fulfilled for suitable geodesic observer
congruences \cite{maluf23,holly}. 
Under this condition, Eq.~\eqref{connThe1} reduces to
\begin{eqnarray}
    u^\lambda \nabla_\lambda{\mathcal{F}^{\hat\alpha}}_{\mu\nu}=\nabla_\mu u^\lambda {\mathcal{F}^{\hat\alpha}}_{\nu\lambda}-\nabla_\nu u^\lambda {\mathcal{F}^{\hat\alpha}}_{\mu\lambda} \, ,
    \label{connThe2}
\end{eqnarray} 
showing that ${\mathcal{F}^{\hat\alpha}}_{\mu\nu}$ is Lie-transported by the vector field $u^\lambda$. 

In analogy with ideal magnetohydrodynamics treatments \cite{pegoraro12}, we introduce a spacelike four-vector $d\ell^\mu=x'^\mu-x^\mu$, which measures the separation between two neighboring events lying on distinct integral curves of the vector field $u^\mu$. Simultaneity between these events is established when $d\ell^0=0$. As discussed in Refs.~\cite{magnetoconnection1,magnetoconnection2} for magnetohydrodynamics in curved spacetime, this separation vector evolves according to 
\begin{eqnarray}
    u^\lambda \nabla_\lambda d\ell^\mu=d\ell^\lambda \nabla_\lambda u^\mu\, ,
    \label{eq:separation}
\end{eqnarray} 
which describes how neighboring curves generated by $u^\mu$ deform relative to one another. In this formulation, $u^\mu$ corresponds to the transport velocity associated with the ideal Ohm-type condition satisfied by the gravitational field.

From the evolution equations for ${\mathcal{F}^{\hat\alpha}}_{\mu\nu}$ and $d\ell^\mu$, we determine the dynamics of the contraction $d\ell^\mu {\mathcal{F}^{\hat\alpha}}_{\mu\nu}$ along the integral curves of the vector field $u^\mu$. Using Eqs.~\eqref{connThe2} and \eqref{eq:separation}, it follows that
\begin{eqnarray}
    u^\lambda \nabla_\lambda\left(d\ell^\mu{\mathcal{F}^{\hat\alpha}}_{\mu\nu}\right)&=&d\ell^\lambda \nabla_\lambda u^\mu{\mathcal{F}^{\hat\alpha}}_{\mu\nu}+d\ell^\mu\nabla_\mu u^\lambda {\mathcal{F}^{\hat\alpha}}_{\nu\lambda}\nonumber\\
    &&-d\ell^\mu\nabla_\nu u^\lambda {\mathcal{F}^{\hat\alpha}}_{\mu\lambda} \, .
\end{eqnarray}
Exploiting the antisymmetry of ${\mathcal{F}^{\hat\alpha}}_{\mu\nu}$ in its spacetime indices, we obtain the ``gravitational field-connection equation'' 
\begin{eqnarray}
    u^\lambda \nabla_\lambda\left(d\ell^\mu{\mathcal{F}^{\hat\alpha}}_{\mu\nu}\right) =-\left(\nabla_\nu u^\lambda\right) \left(d\ell^\mu {\mathcal{F}^{\hat\alpha}}_{\mu\lambda}\right) \, , 
     \label{connThe4}
\end{eqnarray} 
which implies that if 
\begin{eqnarray}
     d\ell^\mu {\mathcal{F}^{\hat\alpha}}_{\mu\nu}=0 \, 
     \label{connThe5}
 \end{eqnarray} 
holds initially, then Eq.~\eqref{connThe4} guarantees that $u^\lambda \nabla_\lambda\left(d\ell^\mu{\mathcal{F}^{\hat\alpha}}_{\mu\nu}\right) =0$ for all times (regularity properties of the vector field $u^\mu$ are assumed). Thus, once the contraction $d\ell^\mu {\mathcal{F}^{\hat\alpha}}_{\mu\nu}$ vanishes, it remains zero throughout the spacetime evolution. This implies the existence of gravitational two-surfaces that preserve their topology during the spacetime dynamics. 

The preservation of gravitational two-surfaces under spacetime dynamics mirrors the conservation of electromagnetic two-surfaces in relativistic magnetohydrodynamics~\cite{Carter79,Gralla14,pegoraro16}. In the present Letter, however, these surfaces are defined by a tetrad-valued gravitational field strength rather than by the electromagnetic field. While the underlying mathematical structure is analogous to that encountered in relativistic magnetohydrodynamics, the physical content and dynamical origin of the two constructions are fundamentally different. 
Accordingly, Eq.~\eqref{connThe4} constitutes the gravitational analog of the connection equation in dissipationless magnetohydrodynamics \cite{pegoraro12,AC15,magnetoconnection1,magnetoconnection2}. The condition \eqref{connThe5} identifies the directions along which the gravitational field strength has vanishing contraction with the separation vector, so that the gravitational field is ``frozen-in'' to the transport defined by $u^\mu$.

To express the gravitational field connectivity in terms of ``gravitational electric'' and ``gravitational magnetic'' fields, we make use of the foliation defined by the timelike unit vector $n_\mu$, which fulfills the normalization condition $n_\mu n^\mu=-1$ \cite{ADM62,MTW}. The metric decomposes as $g_{\mu\nu}=\gamma_{\mu\nu}-n_\mu n_\nu$, where $\gamma_{\mu\nu}$ is the induced three-metric on the spacelike hypersurfaces of constant time. Accordingly, the ``gravitational electric'' and ``gravitational magnetic'' fields defined in Eqs. (\ref{eq:E_tensor}) and (\ref{eq:B_tensor}) are purely spatial, obeying $n_\mu \mathcal{E}^{\hat\alpha\mu}=0=n_\mu \mathcal{B}^{\hat\alpha\mu}$.
In terms of these quantities, the tensor ${\mathcal{F}^{\hat\alpha}}_{\mu\nu}$ and its dual admit the standard decomposition 
\begin{eqnarray}
{\mathcal{F}^{\hat\alpha\mu\nu}}&=&\mathcal{E}^{\hat\alpha\mu} n^\nu-\mathcal{E}^{\hat\alpha\nu} n^\mu-\epsilon^{\mu\nu\rho\lambda}{\mathcal{B}^{\hat\alpha}}_{\rho}n_\lambda\, , \label{expanF1} \\
{^*\mathcal{F}}^{\hat\alpha\mu\nu}&=&\mathcal{B}^{\hat\alpha\mu} n^\nu-\mathcal{B}^{\hat\alpha\nu} n^\mu+\epsilon^{\mu\nu\rho\lambda}{\mathcal{E}^{\hat\alpha}}_{\rho}n_\lambda\, ,
\label{expanF2}
\end{eqnarray}
where $\epsilon^{\mu\nu\rho\lambda}$ is the completely antisymmetric Levi-Civita tensor. 

With this decomposition, condition \eqref{connThe0} can be written explicitly in terms of ``gravitational electric'' and ``gravitational magnetic'' fields.  Condition \eqref{connThe0}
then implies $(u^\mu {\mathcal{E}^{\hat\alpha}}_\mu)n_\nu-\Gamma {\mathcal{E}^{\hat\alpha}}_\nu-u^\mu \epsilon_{\mu\nu\rho\lambda}\mathcal{B}^{\hat\alpha \rho}n^\lambda=0$, where $\Gamma\equiv n_\mu u^\mu$. From this relation we obtain the ideal condition for the gravitational field
\begin{eqnarray}
{\mathcal{E}^{\hat\alpha}}_\mu&=&\frac{1}{\Gamma}\epsilon_{\mu\nu\rho\lambda}u^\nu \mathcal{B}^{\hat\alpha \rho}n^\lambda \, ,
\label{idealC}
\end{eqnarray} 
which is the gravitational analog of the ideal Ohm's law familiar from magnetohydrodynamics. Equation~\eqref{idealC} implies the orthogonality relations $u^\mu {\mathcal{E}^{\hat\alpha}}_\mu=0$ and ${\mathcal{B}_{\hat\alpha}}^\mu {\mathcal{E}^{\hat\alpha}}_\mu=0$. Using these relations, one also obtains ${\mathcal{B}_{\hat\alpha\mu}} \mathcal{F}^{\hat\alpha\mu\nu}= {\mathcal{B}_{\hat\alpha\mu}} \mathcal{E}^{\hat\alpha\mu} n^\nu-\mathcal{E}^{\hat\alpha\nu} {\mathcal{B}_{\hat\alpha\mu}}n^\mu-\epsilon^{\mu\nu\rho\lambda}{\mathcal{B}_{\hat\alpha\mu}}{\mathcal{B}^{\hat\alpha}}_{\rho}n_\lambda=0$. 
The ideal condition~\eqref{idealC} further identifies the velocity field that transports the gravitational field,  
\begin{equation}
u^\mu = \Gamma n^\mu + \frac{\Gamma}{{\mathcal{B}^{\hat\alpha}}_\lambda {\mathcal{B}_{\hat\alpha}}^{\lambda}}\,
\epsilon^{\mu\nu\rho\lambda}\,
\mathcal{E}^{\hat\alpha}{}_\nu\,
\mathcal{B}_{\hat\alpha\rho}\,
n_\lambda \, .
\end{equation}
Under transport by this velocity field, gravitational field connectivity is preserved.

With the ideal condition and transport velocity in hand, we now connect these results to the gravitational field-connection constraint. Substituting Eq.~\eqref{expanF1} into the connection condition \eqref{connThe5}, and using the fact that the separation vector satisfies $n_\mu d\ell^\mu =0$, considering $d\ell^0=0$, we obtain
$d\ell_\mu \mathcal{E}^{\hat\alpha\mu} n^\nu -d\ell_\mu \epsilon^{\mu\nu\rho\lambda}{\mathcal{B}^{\hat\alpha}}_{\rho}n_\lambda=0$. Projecting this relation onto the time foliation, we obtain 
\begin{equation}
d\ell_\mu \mathcal{E}^{\hat\alpha\mu}=0 \, ,
\label{eqn:conn5}
\end{equation} 
\begin{equation}
\epsilon_{\mu\nu\rho\lambda}d\ell^\mu {\mathcal{B}^{\hat\alpha\rho}}n^\lambda=0\, . 
\label{eqn:conn6}
\end{equation} 
Equation \eqref{eqn:conn5} shows that the ``gravitational electric'' field $\mathcal{E}^{\hat\alpha\mu}$ has no component along the separation vector $d\ell_\mu$. Equation \eqref{eqn:conn6}, by contrast, shows that the ``gravitational magnetic'' field ${\mathcal{B}^{\hat\alpha\mu}}$ is everywhere tangent to the two-surface generated by $u^\mu$ and $d\ell^\mu$. Therefore, the ``gravitational magnetic'' field lines are ``frozen-in'' to the vector field $u^\mu$, completing the correspondence with ideal magnetohydrodynamics \cite{Newcomb}.
As discussed by Pegoraro for ideal plasmas \cite{pegoraro12}, the simultaneity condition $d\ell^0=0$ does not affect the generality of this conclusion, since if $d\ell^0 \neq 0$, one can always restore simultaneity by performing the transformation 
\begin{equation}
d\ell^\mu \rightarrow d\ell'^\mu=d\ell^\mu-u^\mu d\lambda \, ,
\end{equation} 
with $\lambda$ an arbitrary scalar function, such that $d\ell'^0=0$. 
This transformation leaves the gravitational field-connection equation \eqref{connThe4} unaltered.

The connection condition for the gravitational field-strength tensor ${\mathcal{F}^{\hat\alpha}}_{\mu\nu}$ naturally extends to the dual Nester-Witten form. Writing ${^*{\cal U}_{\hat\alpha}}^{\mu\nu}$ explicitly in terms of ${\mathcal{F}^{\hat\alpha}}_{\mu\nu}$ gives 
\begin{equation} \label{NW_calU}
    {^*{\cal U}_{\hat\alpha}}^{\hat\mu\hat\nu}
    = \frac{1}{2} \left( {\mathcal{F}^{{\hat \mu} {\hat \nu}}}_{\hat \alpha} - {\mathcal{F}^{{\hat \nu} {\hat \mu}}}_{\hat \alpha} - {\mathcal{F}_{\hat \alpha}}^{{\hat \mu} {\hat \nu}} \right) + {\delta^{\hat \nu}}_{\hat \alpha} 
    {\mathcal{F}_{\hat \beta}}^{{\hat \mu} {\hat \beta}} - {\delta^{\hat \mu}}_{\hat \alpha}  {\mathcal{F}_{\hat \beta}}^{{\hat \nu} {\hat \beta}} ,
\end{equation}
from which one readily obtains ${^*{\cal U}_{\hat\alpha}}^{\hat\alpha\hat\nu}=-{\mathcal{F}_{\hat\alpha}}^{\hat\alpha\hat\nu}$. If the connection condition $d\ell_{\hat\beta} {\mathcal{F}_{\hat\alpha}}^{\hat\beta\hat\nu}=0$ holds, we also have $d\ell_{\hat\beta} {\mathcal{F}_{\hat\alpha}}^{\hat\beta\hat\alpha}=0$, and therefore $d\ell_{\hat\beta} {^*{\cal U}_{\hat\alpha}}^{\hat\beta\hat\alpha}=0$. When the connection condition \eqref{connThe5} is fulfilled, using the above relations in Eq.~\eqref{NW_calU} yields the constraint 
\begin{equation}
d\ell_{\hat\nu}\left({^*{\cal U}_{\hat\alpha}}^{\hat\nu\hat\beta}+ {^*{\cal U}^{\hat\nu\hat\beta}}_{\hat\alpha}+\delta^{\hat\nu\hat\beta} \, {^*{\cal U}^{\hat\mu}}_{\hat\mu\hat\alpha}\right)=0 \, ,
    \label{connU}
\end{equation} 
which specifies how the connection property of ${\mathcal{F}^{\hat\alpha}}_{\mu\nu}$ is inherited by the dual Nester-Witten form. Eq.~\eqref{connU} shows that the quantity obeying the analogous connection (frozen-in) condition is not ${^*{\cal U}_{\hat\alpha}}^{\mu\nu}$ alone, but the combination ${^*{\cal U}_{\hat\alpha}}^{\hat\nu\hat\beta}+ {^*{\cal U}^{\hat\nu\hat\beta}}_{\hat\alpha}+\delta^{\hat\nu\hat\beta} \, {^*{\cal U}^{\hat\mu}}_{\hat\mu\hat\alpha}$. Moreover, by the same reasoning, the condition $u_{\hat\nu}({^*{\cal U}_{\hat\alpha}}^{\hat\nu\hat\beta}+ {^*{\cal U}^{\hat\nu\hat\beta}}_{\hat\alpha}+\delta^{\hat\nu\hat\beta} \, {^*{\cal U}^{\hat\mu}}_{\hat\mu\hat\alpha})=0$ is also fulfilled.

To complete the picture, we  consider how the tetrad-projected Einstein tensor $G_{\hat\alpha\nu}$, as given by the Sparling-Nester-Witten relation \eqref{eqn:G_NW}, evolves under transport by the vector field $u^\mu$. Contracting $G_{\hat\alpha\nu}$ with the separation vector $d\ell^\nu$, and applying the directional covariant derivative $u^\lambda \nabla_\lambda$ while using the transport law \eqref{eq:separation}, yields
\begin{equation}
\begin{aligned}
u^\lambda \nabla_\lambda\!\left(d\ell^\nu G_{\hat\alpha\nu}\right)
&=
-\left(\nabla_{\hat\alpha} u^\lambda\right)\, d\ell^\nu G_{\lambda \nu}
+ S_{\hat\alpha} \, ,
\end{aligned}
\label{eq:G_connection_like_noQ}
\end{equation}
with the source term $S_{\hat\alpha}$ given by 
\begin{equation}
\begin{aligned}
S_{\hat\alpha}
&=
\left(d\ell^\mu \nabla_\mu u^\nu\right) G_{\hat\alpha\nu}
+ d\ell^\nu\, u^\lambda \nabla_\lambda G_{\hat\alpha\nu}
+ \left(\nabla_{\hat\alpha} u^\lambda\right)\, d\ell^\nu G_{\lambda\nu} \, .
\end{aligned}
\label{eq:G_source_noQ}
\end{equation} 
From Eq.~\eqref{eqn:G_NW} one sees that in general, the terms entering $S_{\hat\alpha}$ do not cancel. Therefore, the contraction $d\ell^\nu G_{\hat\alpha\nu}$ does not obey a strict analog of the frozen-in constraint. Nevertheless, one may still identify special directions $d\ell^\nu$ for which the source term vanishes. In those directions the remaining purely kinematic term survives, but the curvature self-interaction drops out. The required condition is 
\begin{equation}
d\ell^\nu [ (\nabla_{\hat\alpha} u^\lambda)\, G_{\nu\lambda}+ (\nabla_\nu u^\lambda)\, G_{\hat\alpha\lambda}
       + u^\lambda\nabla_\lambda G_{\hat\alpha\nu}\,] = 0 \, ,
\label{eq:S_zero_condition}
\end{equation} 
which selects those spacetime directions for which the deformation associated with $u^\mu$ and the evolution of the projected Einstein tensor balance when contracted with the separation vector.
In these directions the source term disappears even though neither $\nabla_\nu u^\lambda$ nor $u^\sigma\nabla_\sigma G_{\hat\alpha\nu}$ is separately zero. Consequently, the transport contribution appearing in Eq.~\eqref{eq:G_connection_like_noQ} remains the only term governing the evolution of $d\ell^\nu G_{\hat\alpha\nu}$ along such directions.

\textit{``Gravitational Magnetic'' Flux Conservation.}---
The Maxwell-type rewriting of the Einstein field equations leads directly to a flux-conservation law for the ``gravitational magnetic'' field, in close analogy with the magnetic frozen-in flux (Alfv{\'e}n) theorem of ideal magnetohydrodynamics. This follows straightforwardly from expressing the homogeneous equation \eqref{homogeneous} in vector form. Writing the timelike unit normal vector as $n_\mu=(-\alpha, 0,0,0)$ and $n^\mu=(1/\alpha,-\beta^i/\alpha)$, where $\alpha$ is the lapse function and $\boldsymbol{\beta}$ is the shift vector \cite{ADM62,MTW}, the constraint \eqref{homogeneous} takes the vector form 
\begin{equation}
    \frac{\partial}{\partial t}\left(\sqrt{\gamma}\, {\boldsymbol{\mathcal{B}}}^{\hat\alpha} \right)+\nabla\times\left(\alpha^2{\boldsymbol{\mathcal{E}}}^{\hat\alpha}-\boldsymbol{\beta}\times{\boldsymbol{\mathcal{B}}}^{\hat\alpha}\right)=0 \, ,
    \label{homogeneous_vector}
\end{equation}
with $\sqrt{\gamma} =\sqrt{-g}/\alpha$. Here and in the following, the vector product is defined using the spacetime Levi-Civita tensor. The ideal Ohm-type condition \eqref{idealC} can likewise be written in vector form as 
\begin{equation} {\boldsymbol{\mathcal{E}}}^{\hat\alpha}=\frac{1}{\alpha^2}\left( \boldsymbol{\beta}-{\boldsymbol{v}}\right)\times{\boldsymbol{\mathcal{B}}}^{\hat\alpha} \, ,
    \label{ideal3delectric}
\end{equation}
using $\Gamma=-\alpha\, u^0$ and $v^i = \alpha u^i/\Gamma$ for the spatial components. 
Equation~\eqref{ideal3delectric} then gives the velocity 
\begin{eqnarray}
    {\boldsymbol{v}}=\frac{1}{\boldsymbol{\mathcal{B}}^{\hat\alpha}\cdot\boldsymbol{\mathcal{B}}_{\hat\alpha}}\left({\alpha^2}{\boldsymbol{\mathcal{E}}}^{\hat\alpha}- \boldsymbol{\beta}\times{\boldsymbol{\mathcal{B}}}^{\hat\alpha}  \right)\times{\boldsymbol{\mathcal{B}}}_{\hat\alpha}  \, .
\end{eqnarray}

Combining Eq.~\eqref{homogeneous_vector} with Eq.~\eqref{ideal3delectric} yields the gravitational analog of the ideal-magnetohydrodynamics induction equation,
\begin{equation}
    \frac{\partial}{\partial t}\left(\sqrt{\gamma}\, {\boldsymbol{\mathcal{B}}}^{\hat\alpha} \right)-\nabla\times\left({\boldsymbol{v}}\times{\boldsymbol{\mathcal{B}}}^{\hat\alpha}\right)=0 \, .
    \label{inductionequation}
\end{equation}
This equation implies that the flux of the ``gravitational magnetic'' field through a comoving surface, 
\begin{eqnarray}
    \Phi=\int ds \, \sqrt{\gamma} \; {\boldsymbol{\mathcal{B}}}^{\hat \alpha}\cdot {\boldsymbol{s}}_{\hat\alpha} \,  ,
    \label{Magnefluxes}
\end{eqnarray}
is conserved, $d\Phi/dt=0$, where ${\boldsymbol{s}}_{\hat\alpha}$ is the vector normal to the comoving surface.

\textit{Gravitational Helicity Conservation.}---
The ideal condition \eqref{idealC} implies the conservation of a gravitational analog of magnetic helicity. In this case, the helicity current is defined as
\begin{equation}
    \mathcal{K}^\mu={^*\mathcal{F}}^{\hat\alpha\mu\nu} \mathcal{A}_{\hat\alpha\nu }\, ,
\label{eq:helicity}
\end{equation} 
which is the natural generalization of the electromagnetic helicity obtained by replacing the electromagnetic field-strength tensor with the gravitational field-strength tensor ${\mathcal{F}^{\hat\alpha}}_{\mu\nu}$.
The covariant divergence of the helicity current takes the form $\nabla_\mu \mathcal{K}^\mu={^*\mathcal{F}}^{\hat\alpha\mu\nu} {\mathcal{F}}_{\hat\alpha\mu\nu}/2$. Evaluating the right-hand side using Eq.~\eqref{expanF2} gives
\begin{equation}
  \nabla_\mu \mathcal{K}^\mu=-2{\mathcal{B}_{\hat\alpha}}^\mu {\mathcal{E}^{\hat\alpha}}_\mu=0 \, ,
  \label{conservationhelicityK}
\end{equation} 
where the vanishing follows from the ideal condition \eqref{idealC}. Thus, the gravitational helicity is conserved. 
In the $3+1$ decomposition, the temporal component of the helicity current is $\mathcal{K}^0 = {\boldsymbol{\mathcal{A}}}_{\hat\alpha} \cdot {\boldsymbol{\mathcal{B}}}^{\hat\alpha}/\alpha $, so that the total helicity on a spatial hypersurface $\Sigma_t$ is
\begin{equation}
    \mathcal{H} = \int_{\Sigma_t} d^{3}x \, \alpha \sqrt{\gamma} \, \mathcal{K}^0 \, .
\end{equation}
When the ``gravitational magnetic'' field is organized into thin flux tubes with central field lines $\Gamma_i$ and associated fluxes \eqref{Magnefluxes} through cross sections $S_i$ of the tubes, $\mathcal{H}$ admits the standard topological interpretation 
\begin{equation}
\mathcal{H}
= \sum_i (\mathrm{Tw}_i + \mathrm{Wr}_i)\,\Phi_i^2
+ 2\sum_{i<j}
\mathrm{Lk}(\Gamma_i,\Gamma_j)\,\Phi_i\Phi_j \, ,
\end{equation}
where $\mathrm{Tw}_i$ and $\mathrm{Wr}_i$ are the twist and writhe of tube $i$, and $\mathrm{Lk}(\Gamma_i,\Gamma_j)$ is the Gauss linking number between the ``gravitational magnetic'' field lines $\Gamma_i$ and $\Gamma_j$ \cite{ArnoldKhesin}. The frozen-in property guarantees that the fluxes $\Phi_i$ and the linking numbers $\mathrm{Lk}(\Gamma_i,\Gamma_j)$ are preserved, so that $\mathcal{H}$ remains invariant. This makes explicit that gravitational helicity conservation encodes the preservation of field-line linkage.

\textit{Conclusions.}---
In this Letter we have shown that the gravitational field admits frozen-in connectivity structures—two-surfaces and associated field lines whose topology is preserved by the Einstein equations under an ideal Ohm-type condition. The gravitational field-connection equation, Eq.~\eqref{connThe4}, together with the condition $d\ell^\mu {\mathcal{F}^{\hat\alpha}}_{\mu\nu}=0$, ensures that the two-surfaces generated by the ``gravitational magnetic'' field $\mathcal{B}^{\hat{\alpha}\mu}$ and the velocity field $u^\mu$ maintain their connectivity throughout the spacetime evolution. These two-surfaces can be interpreted in terms of ``gravitational magnetic'' field lines once an appropriate time-resetting projection is applied to compensate for the loss of simultaneity between spatially separated events. 

Within the same framework, the Bianchi-type identity for the dual gravitational field strength tensor leads directly to conservation of a ``gravitational magnetic'' flux $\Phi$, providing a gravitational analog of the frozen-in flux theorem of ideal magnetohydrodynamics. We further identify a conserved gravitational helicity $\mathcal{H}$, whose invariance is guaranteed by the ideal Ohm-type condition and which admits a direct topological interpretation in terms of twist, writhe, and mutual linkage of ``gravitational magnetic'' field lines. 

The conservation of gravitational field connections, ``gravitational magnetic'' flux, and gravitational helicity, imposes stringent topological constraints on the spacetime evolution by forbidding transitions between different topological configurations of the ``gravitational magnetic'' worldsheets and field lines. The preservation of the gravitational field topology can be directly linked to the geometric structures that spacetime develops during nonlinear evolution, providing a framework for understanding how such structures arise, persist, and interact in dynamical spacetimes. The topological properties of the gravitational field therefore provide an organizing principle underlying the nonlinear dynamics of spacetime.

\begin{acknowledgments}
F.A.A. acknowledges support from FONDECYT grant No. 1230094. 
M.A.W. acknowledges support from the FONDECYT postdoctoral grant No. 3240441.
L.C. acknowledges support from NSF grant PHY-2308944, NASA ATP award 80NSSC22K0667, and NASA ATP award 80NSSC24K1230.
\end{acknowledgments}


\vspace{-0.1cm}

\begin{thebibliography}{}

$\,$

\bibitem{Lehner01} L. Lehner, Classical Quantum Gravity {\bf 18}, R25 (2001).

\bibitem{Alcubierre08} M. Alcubierre, {\it Introduction to 3+1 Numerical Relativity} (Oxford University Press, Oxford, 2008).

\bibitem{BS10} T. W. Baumgarte and S. L. Shapiro, {\it Numerical Relativity: Solving Einstein’s Equations on the Computer} (Cambridge University Press, Cambridge, 2010).

\bibitem{Pretorius2005} F. Pretorius, Phys. Rev. Lett. {\bf 95}, 121101 (2005).

\bibitem{Campanelli2006} M. Campanelli, C. O. Lousto, P. Marronetti, and Y. Zlochower, Phys. Rev. Lett. {\bf 96}, 111101 (2006).

\bibitem{Baker2006} J. G. Baker, J. Centrella, D.-I. Choi, M. Koppitz, and J. van Meter, Phys. Rev. Lett. {\bf 96}, 111102 (2006).

\bibitem{Abbott2016} B. P. Abbott {\it et al}. (LIGO Scientific Collaboration and Virgo Collaboration), Phys. Rev. Lett. {\bf 116}, 061102 (2016).

\bibitem{Abbott2017} B. P. Abbott {\it et al}. (LIGO Scientific Collaboration and Virgo Collaboration), Phys. Rev. Lett. {\bf 119}, 161101 (2017).

\bibitem{Wheeler63} J. A. Wheeler, {\it Geometrodynamics} (Academic Press, New York, 1963).

\bibitem{Wheeler64} J. A. Wheeler, in {\it Relativity, Groups and Topology (Les Houches Summer School, 1963)}, edited by B. DeWitt and C. DeWitt (Gordon and Breach, New York, 1964), p. 325.

\bibitem{MTW} C. W. Misner, K. S. Thorne, and J. A. Wheeler, {\it Gravitation} (Freeman, San Francisco, 1973).

\bibitem{Thorne12} K. S. Thorne, Science {\bf 337}, 536 (2012).

\bibitem{MaarBass98} R. Maartens and B. A. Bassett, Classical Quantum Gravity {\bf 15}, 705 (1998).

\bibitem{Owen11} R. Owen, J. Brink, Y. Chen, J. D. Kaplan, G. Lovelace, K. D. Matthews, D. A. Nichols, M. A. Scheel, F. Zhang, A. Zimmerman, and K. S. Thorne, Phys. Rev. Lett. {\bf 106}, 151101 (2011).

\bibitem{Nichols11} D. A. Nichols, R. Owen, F. Zhang, A. Zimmerman, J. Brink, Y. Chen, J. D. Kaplan, G. Lovelace, K. D. Matthews, M. A. Scheel, and K. S. Thorne, Phys. Rev. D {\bf 84}, 104028 (2011).

\bibitem{Nichols12} D. A. Nichols, A. Zimmerman, Y. Chen, G. Lovelace, K. D. Matthews, R. Owen, F. Zhang, and K. S. Thorne, Phys. Rev. D {\bf 86}, 104028 (2012).

\bibitem{TB17} K. S. Thorne and R. D. Blandford, {\it Modern Classical Physics} (Princeton University Press, Princeton, 2017).

\bibitem{boye}  S. Boyeneni, J. Wu and E. R. Most, Phys. Rev. Lett. {\bf 135}, 101401 (2025).

\bibitem{holly} H. Krynicki, J. Wu and E. R. Most, arXiv:2509.19769.

\bibitem{maluf23} J. W. Maluf, F. L. Carneiro, S. C. Ulhoa, and J. F. d. Rocha-Neto, Ann. Phys. 535, 2300241 (2023)

\bibitem{olivares} H. Olivares, I. M. Peshkoz, E. R. Most, F. M. Guercilena and L. J. Papenfort, Phys. Rev. D {\bf 105}, 124038 (2022).

\bibitem{Peshkov} I. Peshkov, H. Olivares, E. Romenski, Phys. Rev. D {\bf 112}, 084070 (2025).

\bibitem{Lichnerowicz67} A. Lichnerowicz, {\it Relativistic Hydrodynamics and Magnetohydrodynamics} (Benjamin, New York, 1967).

\bibitem{Anile05} A. M. Anile, {\it Relativistic Fluids and Magneto-Fluids} (Cambridge University Press, Cambridge, England, 1989).

\bibitem{Nester81} J. M. Nester, Phys. Lett. {\bf 83A}, 241 (1981).

\bibitem{Witten81} E. Witten, Commun. Math. Phys. {\bf 80}, 381 (1981).

\bibitem{Frauendiener90} J. Frauendiener, Gen. Relativ. Gravit. 22, 1423 (1990).

\bibitem{Szabados92} L. B. Szabados, Classical Quantum Gravity \textbf{9}, 2521 (1992).

\bibitem{LL_V8} L. D. Landau and E. M. Lifshitz, {\it Electrodynamics of Continuous Media, Vol. 8 of Course of Theoretical Physics}, 2nd ed. (Pergamon Press, Oxford, 1984).

\bibitem{pegoraro12} F. Pegoraro, Europhys. Lett {\bf 99}, 35001 (2012). 

\bibitem{magnetoconnection1} F. A. Asenjo and L. Comisso, Phys. Rev. D {\bf 96}, 123004 (2017). 

\bibitem{magnetoconnection2} L. Comisso and F. A. Asenjo, Phys. Rev. D {\bf 102}, 023032 (2020).

\bibitem{Carter79} B. Carter, in \emph{Active Galactic Nuclei}, edited by C. Hazard and S. Mitton (Cambridge University Press, Cambridge, England, 1979), p. 273.

\bibitem{Gralla14} S. E. Gralla and T. Jacobson, Mon. Not. R. Astron. Soc. 445, 2500 (2014).

\bibitem{pegoraro16} F. Pegoraro, J. Plasma Phys. 82, 555820201 (2016)

\bibitem{AC15} F. A. Asenjo and L. Comisso, Phys. Rev. Lett. {\bf 114}, 115003 (2015).

\bibitem{ADM62} R. Arnowitt, S. Deser, and C. W. Misner, in {\it Gravitation: An Introduction to Current Research}, edited by L. Witten (Wiley, New York, 1962), p. 227.

\bibitem{Newcomb} W. A. Newcomb, Ann. Phys. (N.Y.) {\bf 3}, 347 (1958).

\bibitem{ArnoldKhesin} V. I. Arnold and B. A. Khesin, {\it Topological Methods in Hydrodynamics} (Springer-Verlag, Berlin, 1998).

\end{thebibliography}
\end{document}